\documentclass[epj,referee]{svjour}
\usepackage{graphicx}
\usepackage{latexsym}
\usepackage{amssymb}
\usepackage{cases}
\usepackage[compress]{cite}
\usepackage{xcolor}
\usepackage{enumerate}
\usepackage{indentfirst}
\newtheorem{Theorem}{Theorem}

\begin{document}

\title { Average path length for Sierpinski pentagon}

\author{Junhao Peng\inst{1,2}
\thanks{\emph{Email:} pengjh@gzhu.edu.cn}%
\and Guoai Xu\inst{2}
}                     
%
%
\institute{College of Math and Information Science , Guangzhou University , Guangzhou 510006 ,People¡¯s Republic of China. \and  State Key Laboratory of Networking and Switching Technology, Beijing University of Posts and Telecommunications , Beijing 100876 , People¡¯s Republic of China.}
%
\abstract{
In this paper,we investigate diameter and  average path length(APL) for Sierpinski pentagon based on its recursive construction and self-similar structure.We find that the diameter of Sierpinski pentagon is just the shortest path length between  two nodes  of generation $0$.  Deriving and solving the linear homogenous recurrence relation the diameter satisfies,we obtain rigorous  solution for the diameter .We also obtain approximate  solution for APL of Sierpinski pentagon, both diameter and APL  grow approximately as a power-law function of network order $N(t)$,with the exponent equals $\frac{\ln(1+\sqrt{3})}{\ln(5)}$.
Although the solution for APL  is approximate,it is trusted because we have calculated all items of APL accurately except for the compensation( $\Delta_{t}$) of  total distances between non-adjacent branches( $\Lambda_t^{1,3}$) ,which is obtained approximately by least-squares curve fitting.The compensation( $\Delta_{t}$) is only a small part of total distances between non-adjacent branches( $\Lambda_t^{1,3}$) and has little effect on  APL.Further more,using the data obtained by iteration to test the fitting results,we find the  relative error for  $\Delta_{t}$ is less than $10^{-7}$  ,hence the approximate solution for average path length is almost accurate.
\PACS{
     {02.10.Ox }{Combinatorics graph theory}  \and
     {89.75.Hc}{Networks and genealogical trees}  \and
     {06.30.Bp}{Spatial dimensions}
   } 
} 

\maketitle
\section{Introduction}
\label{intro}
Recently,complex networks have attracted a surge of interest from the scientific community \cite{Ref1,Ref2,Ref3,Ref4,Ref5}. Most endeavors were devoted to unveil the structural properties of real network systems ,such as degree distribution\cite{Ref4,Ref5,Ref6} ,degree correlation\cite{Ref7,Ref8},clustering coefficient\cite{Ref9,Ref10},spectral properties\cite{Ref11,Ref12,Ref48,Ref49},diameter\cite{Ref13,Ref14},average path length\cite{Ref15,Ref16},communicability\cite{EEstrada08,EEstrada09},etc.These properties play significant roles in characterizing and understanding complex network systems .\par
Among these structural properties, average path length (APL) ,which is the mean  of the shortest path  lengths between all pairs of vertices,characterizes the small-world behavior commonly observed in  real networks \cite{Ref15,Ref16},it is also related to other structural properties, such as degree distribution \cite{Ref17,Ref18},fractality\cite{Ref19,Ref20},etc. Further more, average path length has an important consequence for dynamical processes taking placing on networks, including disease spreading \cite{Ref15,Ref21,Ref22,Ref23}, routing \cite{Ref24,Ref25,Ref26},  percolation \cite{Ref27} , target search \cite{Ref28,Ref29},and so on .Thus lots of endeavors were devoted to uncovered the APL of different networks,such as Watts-Strogatz model\cite{Ref30}, Barab\'{a}si-Albert network\cite{Ref31},Apollonian network \cite{Ref32},Sierpinski network  \cite{Ref33} and hierarchical scale-free network\cite{Ref34},etc.\par
Sierpinski pentagon belongs to the famous family of Sierpinski objects\cite{Ref35,Ref36},Lots of job was devoted to study the properties of these objects which also include Sierpinski gasket\cite{Ref37,Ref38,Ref39,Ref40,Ref47},Sierpinski carpet\cite{Ref41,Ref42,Ref43}and Sierpinski lattice \cite{Ref44,Ref45} ,etc.As for Sierpinski pentagon, to the best of our knowledge, related research was rarely reported ,and the analytical solution for  average path length has not been addressed.\par
To fill this gap, in this paper, we investigate and obtain approximate solution for average path length . The analytic
method is based on the recursive construction and self-similar structure of Sierpinski pentagon. Our  results show that the average path length increases approximately algebraically with network order .Although the solution for APL  is approximate,it is trusted because we have calculated all items of APL accurately except for the compensation( $\Delta_{t}$) of  total distances between non-adjacent branches( $\Lambda_t^{1,3}$).Further more ,the relative errors for  $\Delta_{t}$ is less than $10^{-7}$ as Sec.\ref{sec:32} shows.\par
In the process of calculating average path length,we also find that the diameter of Sierpinski pentagon is just the shortest path length between  two nodes  of generation $0$ which has been proved in sec.\ref{sec:2}.We derive difference equation  to depict the evolution of diameter ,solving the difference equation ,we gain the rigorous result which shows that the diameter also increases algebraically with network order .
\section{Brief introduction to Sierpinski pentagon}
\label{sec:1}
\begin{figure}
\begin{center}
\includegraphics[scale=1]{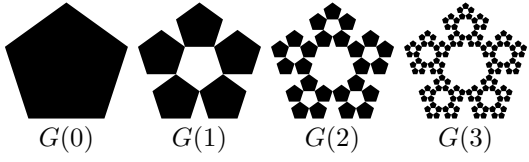}
\caption{Growth process for Sierpinski pentagon from generation $0$ to generation $3$ }
\label{fig:1}       
\end{center}
\end{figure}

Sierpinski pentagon we considered is a fractal which can be constructed iteratively \cite{Ref35,Ref36}. We denote the Sierpinski pentagon after t iterations by $G(t)$ with $t\geq 0 $.Then the fractal is constructed as follows. For $t = 0, G(0)$ is a filled regular pentagon.In order to obtain $G(1)$ ,we divide the regular pentagon $G(0)$  so that $6$ inner pentagons can be drawn out of it, paint all the inner pentagons but the middle one.Apply the same process to the inner pentagons but the middle one, Sierpinski pentagon is the limiting set for this construction.In this paper,the number of iterations for Sierpinski pentagon is called the generation of Sierpinski pentagon .The Sierpinski pentagon for the first four generation is shown in Figure \ref{fig:1}.
According to the construction of Sierpinski pentagon, one can see that at each step t, the total number of edges in the systems increases by a factor of $5$. Thus,  the total number of edges for $G(t)$ is $E_t=5^{t+1}$.We can also find that the total number of nodes which is denoted by $N(t)$ satisfies
$$ N(t)=5\cdot N(t-1)-5$$
Notice $N(0)=5$,we can  obtain
\begin{equation}
N(t)=\frac{3}{4}\cdot 5^{t+1}+\frac{5}{4}
\label{equ1}
\end{equation}  \par
\section{Analytical solution of Diameter }
\label{sec:2}
\begin{figure}
\begin{center}
\includegraphics[scale=0.6]{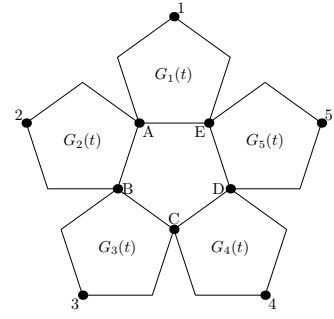}
\caption{Second construction method of
$G(t)$ that highlights self-similarity. The Sierpinski pentagon $G(t+1)$ is composed of five copies of $G(t)$ denoted as  $G_1(t)$,$G_2(t)$,$G_3(t)$,$G_4(t)$,$G_5(t)$.
}
\label{fig:2}
\end{center}
\end{figure}

The diameter of a graph is the maximum of  the shortest path lengths between any pair of nodes,For Sierpinski pentagon,its self-similar structure  allows one to find and calculate diameter analytically.The self-similar structure is obvious from an equivalent network construction method: to obtain $G(t+1)$, one can make five copies of $G(t)$ and join them at the five nodes( i.e. , A,B,C,D and E in Figure \ref{fig:2}). We can see that the  $G(t+1)$ is obtained by the juxtaposition of $5$ copies of $G(t)$ which are  labeled as $G_1(t)$,$G_2(t)$,$G_3(t)$,$G_4(t)$ and $G_5(t)$ ,respectively.\par
We label the five nodes of generation $0$  as $1,2,3,4,5$ and let $d_{ij}(t)$ denotes  the shortest path length from node $i$ to $j$ in $G(t)$,  it is easy to know: $d_{12}(0)=1,d_{12}(1)=4$,$d_{13}(0)=2,d_{13}(1)=5$,and for any $t>1$
\begin{eqnarray}
d_{12}(t)&=&d_{1A}(t-1)+d_{A2}(t-1)=2d_{13}(t-1)
\label{eq2}
\end{eqnarray}
\begin{eqnarray}
d_{13}(t)&=&d_{1A}(t-1)+d_{AB}(t-1)+d_{B3}(t-1) \nonumber \\
 &=&2d_{13}(t-1)+d_{12}(t-1)
\label{eq3}
\end{eqnarray}
It follows that
\begin{equation}
d_{13}(t)=2d_{13}(t-1)+2d_{13}(t-2)
\label{eq4}
\end{equation}
\begin{equation}
d_{12}(t)=2d_{12}(t-1)+2d_{12}(t-2)
\label{eq5}
\end{equation}
Both $d_{12}(t)$ and $d_{13}(t)$ satisfies the same linear homogenous recurrence relation
$$y_t=2y_{t-1}+2y_{t-2}$$
whose general solution\cite{Ref46} is
\begin{equation}
y_t=c_1\cdot \lambda_1^t+c_2\cdot \lambda_2^t
\label{eq6}
\end{equation}
where $\lambda_1, \lambda_2$ is two roots of its characteristic equation $\lambda^2-2\lambda-2=0$,and $c_1, c_2$ is determined by its initial conditions.Sloving the characteristic equation,we have
\begin{eqnarray}
\lambda_1=1+\sqrt{3}, \lambda_2=1-\sqrt{3} \nonumber
\end{eqnarray}
Using the initial conditions $d_{12}(0)=1,d_{12}(1)=4$,$d_{13}(0)=2,d_{13}(1)=5$,we have
\begin{equation}
d_{12}(t)=\frac{1}{2}\cdot (1+\sqrt{3})^{t+1}+\frac{1}{2}\cdot (1-\sqrt{3})^{t+1}
\label{eq7}
\end{equation}
\begin{eqnarray}
d_{13}(t)&=&\frac{2+\sqrt{3}}{2}\cdot (1+\sqrt{3})^t+\frac{2-\sqrt{3}}{2}\cdot (1-\sqrt{3})^t \nonumber \\
&=&\frac{1}{4}\cdot (1+\sqrt{3})^{t+2}+\frac{1}{4}\cdot (1-\sqrt{3})^{t+2}
\label{eq8}
\end{eqnarray} \par
In the infinite system size, i.e.,$ t \rightarrow \infty$
\begin{eqnarray}
d_{13}(t)& \approx &\frac{(1+\sqrt{3})^{t+2}}{4} \nonumber \\
&=&\frac{1+\sqrt{3}}{4}\cdot[\frac{4}{3}(N(t)-\frac{5}{4})]^{\frac{\ln(1+\sqrt{3})}{\ln(5)}} \nonumber \\
& \varpropto & N(t)^{\frac{\ln(1+\sqrt{3})}{\ln(5)}}
\label{eq9}
\end{eqnarray} \par
As the diameter of a graph is the maximum of  the shortest path lengths between any pair of its nodes,we  find that the diameter of Sierpinski pentagon $G(t)$ is just $d_{13}(t)$ which has prooved in Theorem 1.Thus, the diameter grows approximately as a power-law function of network order $N(t)$, with the exponent  is $\frac{\ln(1+\sqrt{3})}{\ln(5)}$.
\begin{Theorem}
For Sierpinski pentagon,let  $L(t)$ denote the diameter of Sierpinski pentagon $G(t)$,thus,
$L(t)=d_{13}(t)$ .
\end{Theorem}
{\bf Proof}:In fact,we want to proof that the inequality
\begin{equation}
d_{ij}(t)\leq d_{13}(t)
\label{equ2}
\end{equation}
 holds for any $t>0$,and any two nodes $i,j$ in $G(t)$.\par
Here we prove the result by mathematical induction .\par
Initial step:For t=0,it is easy to know that the inequality (\ref{equ2}) holds.\par
 Inductive step:Assume there is a $k \geq 0$, such that inequality (\ref{equ2}) holds for $t=k$, we must prove the inequality (\ref{equ2}) holds for $t=k+1$.\par
 For any two nodes $i,j$ of $G(k+1)$,if $i,j$ belong to the same $G(k+1)$ branch which is a copy of $G(k)$,Thus inequality (\ref{equ2}) holds because
 \begin{eqnarray}
 d_{ij}(k+1)=d_{ij}(k)\leq d_{13}(k)<d_{13}(k+1) \nonumber
\end{eqnarray} \par
If $i,j$ belong to two different branches of $G(k+1)$ ,it can be discussed on two cases according the relation of the two different branches.\par
I)If the two  branches is adjacent,by symmetry,we can suppose that $i$ belongs to $G_1(t)$,$j$ belongs to $G_2(t)$,the inequality (\ref{equ2}) holds because
 \begin{eqnarray}
 d_{ij}(k+1)&=&d_{iA}(k+1)+d_{Aj}(k+1) \nonumber \\
 & \leq &d_{1A}(k+1)+d_{A2}(k+1) \nonumber \\
 &<&d_{13}(k+1) \nonumber
\end{eqnarray} \par
II)If the two  branches is not adjacent,by symmetry,we can suppose that $i$ belongs to $G_1(t)$,$j$ belongs to $G_3(t)$,we have
 \begin{eqnarray}
 d_{ij}(k+1)&\leq &d_{iA}(k+1)+d_{AB}(k+1)+d_{Aj}(k+1) \nonumber \\
 & \leq & d_{1A}(k+1)+d_{AB}(k+1)+d_{A2}(k+1)  \nonumber\\
 &=&d_{13}(k+1) \nonumber
\end{eqnarray}
Thus inequality (\ref{equ2}) holds for this case which  finish the proof.
\section{Derivation of Average path length}
\label{sec:3}

We represent all the shortest path lengths of the  Sierpinski pentagon $G(t)$ as a matrix in which the entry $d_{ij}(t)$ is the shortest distance from node $i$ to node $j$, and $d_t$ denotes the average path length (APL) of $G(t)$ which is defined as the mean of $d_{ij}(t)$ over all couples of nodes,thus
\begin{equation}
d_t=\frac{D_t}{N(t)(N(t)-1)/2}
\label{eq11}
\end{equation}
where
\begin{equation}
D_t=\sum_{i,j \in G(t),i\ne j}d_{ij}(t)
\label{eq12}
\end{equation}
denotes the sum of the shortest path length between two nodes over all pairs.\par
Based on the  self-similar structure of $G(t+1)$  As shown in Figure.\ref{fig:2}, it is easy to see that the total distance $D_t$ satisfies \begin{equation}
D_{t+1}=5D_{t}+\Lambda_t
\label{eq13}
\end{equation}
where $\Lambda_t$ ,named the crossing distance in this paper ,denotes the sum over all shortest paths whose end points are not in the same  branch.\par
Let $\Lambda_t^{i,j}$ denotes  the sum of all shortest paths whose endpoints are in $G_i(t)$ and  $G_j(t)$ excluding paths whose end nodes are in the same branch,and $D_t^{i,\alpha}$ denotes the sum of all shortest paths from hub node $\alpha$ (i.e., A,B,C,D and E in figure\ref{fig:2} )to any nodes in  $G_i(t)$ ,that is to say
\begin{eqnarray}
\Lambda_t^{1,2}=\sum_{i \in G_1(t),j\in G_2(t)\atop i,j\ne A}d_{ij}(t)
\label{eq14}
\end{eqnarray}
\begin{eqnarray}
\Lambda_t^{1,3}=\sum_{i \in G_1(t),j\in G_3(t)\atop i\ne A,j\ne B}d_{ij}(t)
\label{eq15}
\end{eqnarray}
\begin{eqnarray}
D_t^{1,C}=\sum_{i \in G_1(t)}d_{iC}(t)
\label{eq16}
\end{eqnarray}
It is easy to know from the self-similar structure of $G(t+1)$
\begin{eqnarray}
\Lambda_t=&\Lambda_t^{1,2}+\Lambda_t^{2,3}+\Lambda_t^{3,4}+\Lambda_t^{4,5}+\Lambda_t^{5,1}  \nonumber \\
&+\Lambda_t^{1,3}+\Lambda_t^{1,4}+\Lambda_t^{2,4}+\Lambda_t^{2,5}+\Lambda_t^{3,5}  \nonumber \\
&-D_t^{1,C}-D_t^{2,D}-D_t^{3,E}-D_t^{4,A}-D_t^{5,B}
\label{eq17}
\end{eqnarray}
The last five terms of Eq. (\ref{eq17}) which want to be subtracted are the items which have been calculated twice .For example,$D_t^{1,C}$ (excluding $d_{AC}(t)$ )is calculated both in $\Lambda_t^{1,3}$ and $\Lambda_t^{1,4}$,and $d_{AC}(t)$ is calculated both in $\Lambda_t^{2,3}$ and $\Lambda_t^{1,4}$.
By symmetry,we have
\begin{eqnarray}
\Lambda_t=5\Lambda_t^{1,2}+5\Lambda_t^{1,3}-5D_t^{1,C}
\label{eq18}
\end{eqnarray}
\subsection{Total distances from one node of generation $0$ to any other nodes}
In this section ,we will calculate a quantity which will be used in calculating $\Lambda_t $ ,the quantity denoted by $S_t$ is the total distances from one node of generation $0$ (labeled by $1,2,3,4,5$ which was shown in Figure.\ref{fig:2} )to  any other nodes of $G(t)$ .It is easy to know that, this  quantity is equal for any nodes of generation $0$,thus
\begin{equation}S_t=\sum_{i \in G(t),i\ne 5}d_{i5}(t)\end{equation} \par
It is easy to know that $S_0=6$. We also find that $S_t$ satisfies the recurrence relations derived as follows,which will help us to obtain the analytical solution for $S_t$ .\par
Note that $G(t+1)$ is obtained by the juxtaposition of $5$ copies of $G(t)$ which are  labeled as $G_1(t)$,$G_2(t)$,$G_3(t)$,$G_4(t)$ and $G_5(t)$ ,respectively,we have
\begin{eqnarray}
S_{t+1}&=&\sum_{i \in G_5(t),i\ne 5}d_{i5}(t+1)+\sum_{i \in G_1(t),i\ne E}d_{i5}(t+1) \nonumber\\
&+&\sum_{i \in G_4(t),i\ne D}d_{i5}(t+1) + \sum_{i \in G_2(t),i\ne A}d_{i5}(t+1)  \nonumber\\
   &+& \sum_{i \in G_3(t),i\ne C}d_{i5}(t+1)-d_{B5}(t+1)
  \label{eq19}
\end{eqnarray}
where
\begin{eqnarray}
\sum_{i \in G_5(t),i\ne 5}d_{i5}(t+1)&=&S_{t} \nonumber\\
\sum_{i \in G_1(t),i\ne E}d_{i5}(t+1)&=&\sum_{i \in G_1(t),i\ne E}(d_{iE}(t+1)+d_{E5}(t+1) )\nonumber\\
    &=&S_{t}+(N(t)-1)d_{13}(t) \nonumber \\
 \sum_{i \in G_4(t),i\ne D}d_{i5}(t+1)&=&S_{t}+(N(t)-1)d_{13}(t) \nonumber   \\
 \sum_{i \in G_2(t),i\ne A}d_{i5}(t+1)&=&S_{t}+(N(t)-1)(d_{12}(t)+d_{13}(t)) \nonumber\\
 \sum_{i \in G_3(t),i\ne C}d_{i5}(t+1)&=&S_{t}+(N(t)-1)(d_{12}(t)+d_{13}(t)) \nonumber
  \end{eqnarray}
 \begin{eqnarray}
  d_{B5}(t)=d_{BA}(t)+d_{AE}(t)+d_{E5}(t)=2d_{12}(t)+d_{13}(t) \nonumber
  \end{eqnarray}
Thus
\begin{eqnarray}
S_t&=&5S_{t-1}+(2N(t-1)-4)d_{12}(t-1)  \nonumber \\
   &+&(4N(t-1)-5)d_{13}(t-1)  \nonumber \\
   &\equiv &5S_{t-1}+f(t-1) \nonumber \\
   &=&5(5S_{t-2}+f(t-2))+f(t-1) \nonumber \\
   &=&5^2S_{t-2}+5f(t-2)+f(t-1) \nonumber \\
   &=& \ldots  \nonumber \\
   &=& 5^tS_{0}+5^{t-1}f(0)+\ldots+5f(t-2)+f(t-1) \nonumber \\
   &=& 5^t\cdot 6+\sum_{i=0}^{t-1}[5^{t-1-i}f(i)]
 \label{eq20}
\end{eqnarray}
with
\begin{eqnarray}
f(t)\equiv(2N(t)-4)d_{12}(t)+(4N(t)-5)d_{13}(t) \nonumber
\end{eqnarray}
It follows from Eqs.(\ref{equ1}),(\ref{eq7}) and (\ref{eq8}) that
\begin{eqnarray}
& &\sum_{i=0}^{t-1}[5^{t-1-i}f(i)] \nonumber \\
&=&\sum_{i=0}^{t-1}5^{t-1-i}[(2N(i)-4)d_{12}(i)+(4N(i)-5)d_{13}(i)]  \nonumber \\
   &=&\sum_{i=0}^{t-1}5^{t-1-i}[(\frac{3}{2}5^{i+1}-\frac{3}{2})d_{12}(i)+3\cdot 5^{i+1}d_{13}(i)]  \nonumber \\
   &=&\frac{3}{2}\sum_{i=0}^{t-1}5^{t}d_{12}(i)-\frac{3}{2}\sum_{i=0}^{t-1}5^{t-1-i}d_{12}(i)+3 \sum_{i=0}^{t-1}5^{t}d_{13}(i)  \nonumber \\
   &=& \frac{3}{2}\cdot 5^{t} \sum_{i=0}^{t-1}[\frac{1}{2}\cdot (1+\sqrt{3})^{i+1}+\frac{1}{2}\cdot (1-\sqrt{3})^{i+1}] \nonumber \\
   &  &-\frac{3}{2}\sum_{i=0}^{t-1}5^{t-1-i}[\frac{1}{2}\cdot (1+\sqrt{3})^{i+1}+\frac{1}{2}\cdot (1-\sqrt{3})^{i+1}]\nonumber \\
   &  &+3\cdot 5^{t}\sum_{i=0}^{t-1}[\frac{1}{4}\cdot (1+\sqrt{3})^{i+2}+\frac{1}{4}\cdot (1-\sqrt{3})^{i+2}]   \nonumber \\
    &=& \frac{3}{2} 5^{t} [\frac{1+\sqrt{3}}{2} \frac{(1+\sqrt{3})^t-1}{\sqrt{3}}+\frac{1-\sqrt{3}}{2} \frac{(1-\sqrt{3})^t-1}{-\sqrt{3}}] \nonumber \\
   &  &-\frac{3}{2}5^{t-1}[\frac{1+\sqrt{3}}{2} \frac{(\frac{1+\sqrt{3}}{5})^t-1}{\frac{1+\sqrt{3}}{5}-1}+\frac{1-\sqrt{3}}{2} \frac{(\frac{1-\sqrt{3}}{5})^t-1}{\frac{1-\sqrt{3}}{5}-1}]\nonumber \\
   &  &+3\cdot 5^{t}[\frac{2+\sqrt{3}}{2} \frac{(1+\sqrt{3})^t-1}{\sqrt{3}}+\frac{2-\sqrt{3}}{2} \frac{(1-\sqrt{3})^t-1}{-\sqrt{3}}]  \nonumber \\
   &=& 5^t[\frac{9+5\sqrt{3}}{4}(1+\sqrt{3})^{t}+\frac{9-5\sqrt{3}}{4}(1-\sqrt{3})^{t}-\frac{69}{13}] \nonumber \\
   &  &+ \frac{21+15\sqrt{3}}{52}(1+\sqrt{3})^{t}+\frac{21-15\sqrt{3}}{52} (1-\sqrt{3})^{t}
 \label{eq21}
\end{eqnarray}
Hence
 \begin{eqnarray}
S_t&=&5^t[\frac{9+5\sqrt{3}}{4}(1+\sqrt{3})^{t}+\frac{9-5\sqrt{3}}{4}(1-\sqrt{3})^{t}+\frac{9}{13}] \nonumber \\
   &+& \frac{21+15\sqrt{3}}{52}(1+\sqrt{3})^{t}+\frac{21-15\sqrt{3}}{52} (1-\sqrt{3})^{t}
 \label{eq22}
\end{eqnarray}
\subsection{Total distances of Non-adjacent branch: $\Lambda_t^{1,3}$ }
\label{sec:32}
Now,we will derive the total distances between branch $G_1(t)$ and $G_3(t)$ which is denoted by $\Lambda_t^{1,3}$. According to the construction shown in Figure.\ref{fig:2},we can find that, for most pairs of nodes $i,j(i \in G_1(t)$,$j \in G_3(t)$), the shortest path pass through node $A$ and $B$ ,hence
\begin{eqnarray}
\Lambda_t^{1,3}&=&\sum_{i \in G_1(t),j\in G_3(t)\atop i\ne A,j\ne B}d_{ij}(t)  \nonumber \\
               &=&\sum_{i \in G_1(t),j\in G_3(t)\atop i\ne A,j\ne B}[d_{iA}(t)+d_{AB}(t)+d_{Bj}(t)]-\Delta_{t} \nonumber \\
               &=&\sum_{i \in G_1(t),j\in G_3(t)\atop i\ne A,j\ne B}d_{iA}(t)+\sum_{i \in G_1(t),j\in G_3(t)\atop i\ne A,j\ne B}d_{AB}(t)   \nonumber \\
                & &+\sum_{i \in G_1(t),j\in G_3(t)\atop i\ne A,j\ne B}d_{Bj}(t)-\Delta_{t} \nonumber \\
               &=&(N(t)-1)\sum_{i \in G_1(t),i\ne A}d_{iA}(t)+(N(t)-1)^2d_{AB}(t)               \nonumber \\
               &  &+(N(t)-1)\sum_{j \in G_2(t),j\ne A}d_{Bj}(t)-\Delta_{t} \nonumber \\
                &=&2(N(t)-1)S_{t}+(N(t)-1)^2d_{12}(t)-\Delta_{t}
\label{eq23}
\end{eqnarray}
with $\Delta_{t}$ to compensate for the overcount of certain pairs whose shortest paths does not pass through $A,B$.while $t$ shows that both branch $G_1(t)$ and $G_3(t)$ are a copy of Sierpinski pentagon  $G(t)$.\par
\begin{figure}
\begin{center}
\includegraphics[scale=0.8]{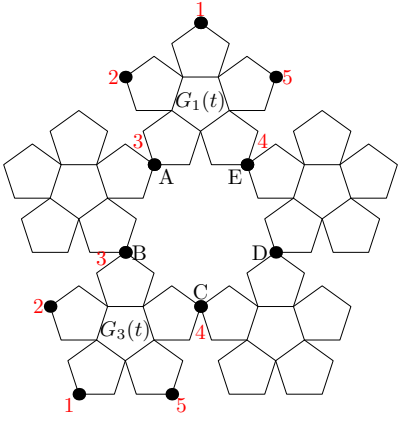}
\caption{The branch $G_1(t)$ and $G_3(t)$ which is looked upon  as a Sierpinski pentagon $G(t)$ with  the five nodes of generation $0$ labeled as $1,2,3,4 $ and $5$.}
\label{fig:3}       
\end{center}
\end{figure}
In order to calculating $\Delta_{t}$,we must know when  overcount occurs and how many it is.For any  pair of nodes $i,j(i \in G_1(t)$,$j \in G_3(t)$), the shortest path can pass through node $A$ and $B$ or pass through $C,D$ and $E$,let $d_{ij}^A(t),d_{ij}^E(t)$ denote the two kinds of  shortest path between $i,j$ in $G(t)$, respectively , we have
\begin{eqnarray}
 d_{ij}^A(t+1)&=&d_{iA}(t+1)+d_{AB}(t+1)+d_{Bj}(t+1) \nonumber \\
             &=&d_{iA}(t+1)+d_{Bj}(t+1)+d_{12}(t)\nonumber \\
 d_{ij}^E(t+1)&=&d_{iE}(t+1)+d_{ED}(t+1)\nonumber \\
             & &+d_{DC}(t+1)+d_{Cj}(t+1)\nonumber \\
              &=&d_{iE}(t+1)+d_{Cj}(t+1)+2d_{12}(t) \nonumber
 \end{eqnarray}
 Thus
 \begin{eqnarray}
 d_{ij}^A(t+1)- d_{ij}^E(t+1)&=&d_{iA}(t+1)-d_{iE}(t+1)\nonumber \\
 &+&d_{Bj}(t+1)-d_{Cj}(t+1)-d_{12}(t) \nonumber
 \end{eqnarray} \par
 If $d_{ij}^A(t+1)- d_{ij}^E(t+1)>0$, overcount occurs ,and it will be added into $\Delta_{t}$,if $d_{ij}^A(t+1)- d_{ij}^E(t+1)\leq 0$,it has no effect on $\Delta_{t}$.\par
 If we look upon  branch $G_1(t)$ and $G_3(t)$ as a Sierpinski pentagon $G(t)$ and label the five nodes of generation $0$ as $1,2,3,4 $ and $5$, which is shown in Figure \ref{fig:3}, we  find that the hub node A ,E ,B,C of $G(t+1)$ is just node $ 3,4$ of $G_1(t)$ and node $ 3,4$ of $G_3(t)$,while $G_1(t)$ and $G_3(t)$ is looked upon as a Sierpinski pentagon $G(t)$ .Thus
  \begin{eqnarray}
 & &d_{ij}^A(t+1)- d_{ij}^E(t+1) \nonumber \\
 &=&d_{i3}(t)-d_{i4}(t)+d_{j3}(t)-d_{j4}(t)-d_{12}(t)
  \label{eq25}
 \end{eqnarray}
which imply that  $d_{ij}^A(t+1)- d_{ij}^E(t+1)$  subjects to  $d_{i3}(t)-d_{i4}(t)$,while $d_{i3}(t)-d_{i4}(t)$ is the distance difference for  node $i$  to node $3$ and $4$ in Sierpinski pentagon $G(t)$. We can calculate $\Delta_{t}$ If we can obtain  $d_{i3}(t)-d_{i4}(t)$ for all nodes $i$ of $G(t)$,which can be solved based on  the recurrence relations $d_{i3}(t)-d_{i4}(t)$ satisfies. \par
 According the construction of $G(t+1)$.we can find the distance difference from $i$ to node $2,3,4$ satisfies recurrence relations which rely on the branch where node $i$ locates.If $i$ is a node of $G_1(t)$
 \begin{eqnarray}
  d_{i3}(t+1)-d_{i4}(t+1)&=&d_{i3}(t)-d_{i4}(t)   \label{eq26}  \\
  d_{i2}(t+1)-d_{i4}(t+1)&=&d_{i3}(t)-d_{i4}(t)-d_{12}(t)
 \end{eqnarray}
If if $i$ is a node of $G_2(t)$
 \begin{eqnarray}
  &&d_{i3}(t+1)-d_{i4}(t+1)=-d_{12}(t)    \\
  &&d_{i2}(t+1)-d_{i4}(t+1) \nonumber \\
  &&=d_{i2}(t)-d_{i4}(t)-d_{12}(t)-d_{13}(t)
  \label{eq29}
 \end{eqnarray}
If $i$ is a node of $G_3(t)$
\begin{eqnarray}
  &&d_{i3}(t+1)-d_{i4}(t+1) \nonumber \\
  &&=d_{i2}(t)-d_{i4}(t)-d_{13}(t)   \\
  &&d_{i2}(t+1)-d_{i4}(t+1) \nonumber \\
  &&=d_{i2}(t)-d_{i4}(t)-d_{13}(t)-d_{12}(t)
   \label{eq28}
 \end{eqnarray}
 If $i$ is a node of $G_4(t)$
 \begin{eqnarray}
  &&d_{i3}(t+1)-d_{i4}(t+1)\nonumber \\
  &&=d_{i2}(t)-d_{i4}(t)+d_{13}(t)   \\
  &&d_{i2}(t+1)-d_{i4}(t+1) \nonumber \\
  &&=d_{i2}(t)-d_{i4}(t)+d_{13}(t)+d_{12}(t)
   \label{eq27}
 \end{eqnarray}
 If $i$ is a node of $G_5(t)$
 \begin{eqnarray}
  &&d_{i3}(t+1)-d_{i4}(t+1)=d_{12}(t)   \\
  &&d_{i2}(t+1)-d_{i4}(t+1)\nonumber \\
  &&=d_{i2}(t)-d_{i3}(t)+d_{12}(t)    \label{eq30}
 \end{eqnarray}\par
Let $\Omega_{3,4}(t),\Omega_{2,4}(t)$ denote the set of  all the  values of $d_{i3}(t)-d_{i4}(t)$ and $d_{i2}(t)-d_{i4}(t)$, respectively. While t=0,$\Omega_{3,4}(0)=\{-1,-1 ,0 ,1,1\}$ ,$\Omega_{2,4}(0)=\{-2, -1, 0 ,1, 2\}$. For  $t>0$,we can obtain $\Omega_{3,4}(t),\Omega_{2,4}(t)$  based on the recurrence relations as Eqs.(\ref{eq26})-(\ref{eq30}) show. \par
 Now we come back to analyze $\Delta_{t}$ which is the sum of all non-negative values of Eq.(\ref{eq25}). It is easy to know that $d_{i3}(t)-d_{i4}(t)$ and $d_{j3}(t)-d_{j4}(t)$ in  Eq.(\ref{eq25}) has the same set of possible values : $\Omega_{3,4}(t)$ which has just been obtained .  We can calculate all the possible values of Eq.(\ref{eq25}),and $\Delta_{t}$ is get by adding all the non-negative values,the results for $t=0 \sim 11$ is shown in Table 1.\par
 \begin{table}
\begin{center}
\caption{\label{tb1}The value of $\Delta_{t}$  for $t=0 \sim 11$ obtained by iteration}
\begin{tabular}{|c|c|c|c|c|c|} \hline
$t$&$\Delta_t$&$t$&$\Delta_t$&$t$&$\Delta_t$  \\ \hline
0&4      &4& 3697330       &8&79817184975658  \\ \hline
1&30     &5&251032868      & 9& $5.45159641\times10^{15}$ \\ \hline
2&1002   &6&171140501308   &10&$3.72349326 \times 10^{17}$ \\ \hline
3&56540  &7&1168705606692  &11&$2.54319349\times 10^{19}$ \\ \hline
\end{tabular}
\end{center}
\end{table}

But with the increasing of $t$ ,it is difficult to calculate $\Delta_t$ by iteration because it is prohibitively time and memory consuming.Substituting Eqs.(\ref{equ1}),(\ref{eq7}) and (\ref{eq22}) into Eq.(\ref{eq23}),we find that the expression for $ \Lambda_t^{1,3}$  satisfies Eq.(\ref{eq31}) if we ignore  $\Delta_{t}$ .We also believe $\Delta_{t}$ can only change $\Lambda_t^{1,3}$ a little ,and  $\Delta_{t}$ can be approximated by Eq.(\ref{eq31}) .
 \begin{eqnarray}
\Phi(t)&=&a_1\cdot5^{2t}(1+\sqrt{3})^{t}+a_2\cdot5^{2t}(1-\sqrt{3})^{t}\nonumber \\
& &+a_3\cdot5^t(1+\sqrt{3})^{t}+a_4\cdot5^t(1-\sqrt{3})^{t} + a_5\cdot(1+\sqrt{3})^{t}\nonumber \\
& &+a_6\cdot(1-\sqrt{3})^{t}+a_7\cdot5^{2t}+a_8\cdot5^t+a_9
 \label{eq31}
\end{eqnarray}
 with the  $9$ coefficients determined by  the actual data shown in Table 1.Using standard software package of MATLAB R2008a ,we  obtain the $9$ coefficients of $\Delta_{t}$ by least-squares curve fitting .Results show the residual is equal to $1.78\times 10^{16}$ and relative error( defined as the absolute error divided by the true value) is also large .If we delete the term $5^{2t}$ in $\Delta_{t}$ whose fitting coefficient ($a_7=-0.0011)$ is small and  conduct least-squares curve fitting again based on data for  $t=0 \sim 7$ in Table 1,the residual is equal to $2.15\times10^{-7}$ which is very small and the $8$ coefficients of $\Delta_{t}$ is:
 \begin{eqnarray}
 &&a_1=0.168524328052979,a_2=-0.0396624946437528 \nonumber \\
 &&a_3=0.935344610079585,a_4=0.951717329999713\nonumber \\
 &&a_5=-4.00947432595951,a_6=-1.49385494978489\nonumber \\
 &&a_8=2.71082171547275,a_9=4.7765837996533
\label{eq32}
\end{eqnarray}
 We  compare the results calculated by fitting curve  and the  data in Table 1 for $t=0 \sim 11$, the relative error is less than $10^{-7}$ which is acceptable and can not be avoided for round-off error,thus our model is trusted.The reason why we don't conduct least-squares curve fitting with more data is the relative error become lager if we use more data for there is huge differences among the data for different $t$.
\subsection{Approximate solution for  average path length}
In this subsection,  we calculate $\Lambda_t^{1,2}$  and $D_t^{1,C} $  first , and then  $\Lambda_t$ and $D_t$ can be obtained from Eqs.(\ref{eq18}) and (\ref{eq13}).Finally,we calculate average path length  from Eq.(\ref{eq11}).\par
According to the  construction of $G(t+1)$,we find
\begin{eqnarray}
\Lambda_t^{1,2}&=&\sum_{i \in G_1(t),j\in G_2(t)\atop i,j\ne A}d_{ij}(t+1)  \nonumber \\
               &=&\sum_{i \in G_1(t),j\in G_2(t)\atop i,j\ne A}[d_{iA}(t+1)+d_{Aj}(t+1)]  \nonumber \\
               &=&2(N(t)-1)S_{t}
\label{eqa1}
\end{eqnarray}
and
\begin{eqnarray}
D_t^{1,C}&=&\sum_{i \in G_1(t)}d_{iC}(t+1) \nonumber \\
         &=&\sum_{i \in G_1(t)}[2d_{12}(t)+min\{d_{iA}(t+1),d_{iE}(t+1)\}] \nonumber \\
         &=&2N(t)d_{12}(t)+\sum_{i \in G(t)}min\{d_{i3}(t),d_{i4}(t)\} \nonumber \\
         &\equiv&2N(t)d_{12}(t)+F(t)
\label{eqa2}
\end{eqnarray}
where
\begin{eqnarray}
F(t)&=&\sum_{i \in G(t)}min\{d_{i3}(t),d_{i4}(t)\} \nonumber \\
    &=&2S_{t-1}+2[N(t-1)-1]d_{13}(t-1) \nonumber \\
    & &+2S_{t-1}-d_{13}(t-1)+F(t-1) \nonumber \\
    & &+[N(t-1)-2][d_{13}(t-1)+d_{12}(t-1)] \nonumber \\
    &=&4S_{t-1}+[3N(t-1)-5]d_{13}(t-1) \nonumber \\
    & &+[N(t-1)-2]d_{12}(t-1)+F(t-1) \nonumber \\
    &=&4S_{t-1}+[3N(t-1)-5]d_{13}(t-1) \nonumber \\
    & &+[N(t-1)-2]d_{12}(t-1) \nonumber \\
    &  &+4S_{t-2}+[3N(t-2)-5]d_{13}(t-2)\nonumber \\
    & &+[N(t-2)-2]d_{12}(t-2)+F(t-2) \nonumber \\
    &=& \cdots \nonumber \\
    &=&4\sum_{k=0}^{t-1}S_k+\sum_{k=0}^{t-1}\{[3N(k)-5]d_{13}(k)\}\nonumber \\
    & &+\sum_{k=0}^{t-1}\{[N(k)-2]d_{12}(k)\}+F(0) \nonumber \\
    &=&4\sum_{k=0}^{t-1}S_k+\sum_{k=0}^{t-1}\{[3N(k)-5]d_{13}(k)\} \nonumber \\
    & &+\sum_{k=0}^{t-1}\{[N(k)-2]d_{12}(k)\}+4  \nonumber \\
    &=&\frac{485\sqrt{3}+792}{472}5^t(1+\sqrt{3})^t+\frac{792-485\sqrt{3}}{472} \nonumber \\
    & &\times5^t(1-\sqrt{3})^t+\frac{48-\sqrt{3}}{312}(1+\sqrt{3})^t \nonumber \\
    & &+\frac{48+\sqrt{3}}{312}(1-\sqrt{3})^t+\frac{9}{13}5^t-\frac{21}{59}
\label{eqa3}
\end{eqnarray}
Thus,the crossing distance
\begin{eqnarray}
\Lambda_t&=&5\Lambda_t^{1,2}+5\Lambda_t^{1,3}-5D_t^{1,C} \nonumber \\
         &=&5\cdot2(N(t)-1)S_{t}+5\cdot[2(N(t)-1)S_{t}+(N(t) \nonumber \\
         & &-1)^2d_{12}(t)-\Delta_{t}]-5\cdot[2N(t)d_{12}(t)+F(t)] \nonumber \\
         &=&20(N(t)-1)S_{t}+5(N(t)-1)^2d_{12}(t) \nonumber \\
         & &-10N(t)d_{12}(t)-5F(t)-5\Delta_{t}
\label{eqa4}
\end{eqnarray}
Substituting Eqs.(\ref{eqa3}),(\ref{equ1}),(\ref{eq7}),(\ref{eq8}),(\ref{eq22}),(\ref{eq32}) into Eq.(\ref{eqa4}),we obtain
%
\begin{eqnarray}
\Lambda_t&=&426.3358\cdot5^{2t}(1+\sqrt{3})^{t}-19.1676\cdot5^{2t}(1-\sqrt{3})^{t}\nonumber\\
& &+29.4512\cdot5^t(1+\sqrt{3})^{t} -0.7143\cdot5^t(1-\sqrt{3})^{t} \nonumber \\
&  &+ 7.1748\cdot(1+\sqrt{3})^{t}+10.6543\cdot(1-\sqrt{3})^{t}\nonumber \\
& & + 51.9230\cdot5^{2t} -13.5541\cdot5^t-22.1032 \nonumber \\
&\equiv&\sum_{k=1}^9{c_kq_k^t}
\label{eqa9}
\end{eqnarray}
with the last line  is  an abbreviation,and $c_k,q_k$
correspond to appropriate expressions shown above. Thus,the total distance
\begin{eqnarray}
D_t&=&5D_{t-1}+\Lambda_{t-1} \nonumber \\
   &=&5^2D_{t-2}+5\Lambda_{t-2}+\Lambda_{t-1} \nonumber \\
   &=&\cdots \nonumber \\
   &=&5^tD_{0}+\sum_{i=0}^{t-1}[5^{t-1-i}\Lambda_{i}] \nonumber \\
   &=&5^tD_{0}+\sum_{i=0}^{t-1}[5^{t-1-i}\sum_{k=1}^9{c_kq_k^i} \nonumber \\
   &=&5^tD_{0}+5^{t-1}\sum_{k=1}^9 {\sum_{i=0}^{t-1} {c_k (\frac{q_k}{5})^i} } \nonumber \\
   &=&5^tD_{0}+5^{t-1}\sum_{k=1}^9 {c_k \frac{1-(\frac{q_k}{5})^t}{1-(\frac{q_k}{5})}}
\label{eqa7}
\end{eqnarray}
It is easy to know that $D_0=15$,substituting  $c_k,q_k$ $(k=1,2,\cdots,9)$ for the appropriate expressions in Eq.(\ref{eqa9}) ,we have
\begin{eqnarray}
D_t&=&\frac{c_1}{5(4+5\sqrt{3})}\cdot5^{2t}(1+\sqrt{3})^{t}\nonumber\\
&&+\frac{c_2}{5(4-5\sqrt{3})}\cdot5^{2t}(1-\sqrt{3})^{t}\nonumber\\
& &+\frac{c_3}{5\sqrt{3}}\cdot5^t(1+\sqrt{3})^{t} -\frac{c_4}{5\sqrt{3}}\cdot5^t(1-\sqrt{3})^{t} \nonumber \\
&  &- \frac{c_5}{4-\sqrt{3}}\cdot(1+\sqrt{3})^{t}- \frac{c_6}{4+\sqrt{3}}\cdot(1-\sqrt{3})^{t} \nonumber \\
& &+ \frac{c_7}{20}\cdot5^{2t}+[15-\frac{c_1}{5(4+5\sqrt{3})}-\frac{c_2}{5(4-5\sqrt{3})}\nonumber \\
& &-\frac{c_3}{5\sqrt{3}}+\frac{c_4}{5\sqrt{3}}+\frac{c_5}{4-\sqrt{3}} +\frac{c_6}{4+\sqrt{3}}-\frac{c_7}{20}\nonumber \\
& &+\frac{c_8t}{5}+\frac{c_9}{4}]\cdot5^t- \frac{c_9}{4} \nonumber \\
&=&6.7350\cdot5^{2t}(1+\sqrt{3})^{t}+0.8226\cdot5^{2t}(1-\sqrt{3})^{t}\nonumber\\
& &+ 3.4007\cdot5^t(1+\sqrt{3})^{t} +0.08248\cdot5^t(1-\sqrt{3})^{t} \nonumber \\
&  &-3.1636\cdot(1+\sqrt{3})^{t}-1.8587\cdot(1-\sqrt{3})^{t}\nonumber \\
& & + 2.5961\cdot5^{2t}+(0.859-2.71t)5^t+5.526
\label{eqa8}
\end{eqnarray}
It follows from Eqs.(\ref{equ1}) and (\ref{eq11})   that
\begin{eqnarray}
d_t&=&\frac{D_t}{\frac{225}{32}5^{2t}+\frac{45}{16}5^{t}+\frac{5}{32}} \nonumber\\
&=&\frac{\frac{D_t}{5^{2t}}}{\frac{225}{32}+\frac{45}{16}5^{-t}+\frac{5}{32}5^{-2t}} \nonumber
\label{eqa10}
\end{eqnarray}
In the infinite system size, i.e.,$ t \rightarrow \infty$
\begin{eqnarray}
d_t&\approx& 0.9579\cdot(1+\sqrt{3})^{t}+0.1170\cdot(1-\sqrt{3})^{t} \nonumber\\
& &+0.4836(\frac{1+\sqrt{3}}{5})^t+0.0117(\frac{1-\sqrt{3}}{5})^t+0.3692 \nonumber\\
&\approx&0.9579\cdot(1+\sqrt{3})^{t} \nonumber\\
 &=&\frac{0.9579}{1+\sqrt{3}}\cdot[\frac{4}{3}(N(t)-\frac{5}{4})]^{\frac{\ln(1+\sqrt{3})}{\ln(5)}}\nonumber\\
 &\varpropto& N(t)^{\frac{\ln(1+\sqrt{3})}{\ln(5)}}
\label{eqa19}
\end{eqnarray} \par
which implies that  APL grows approximately as a power-law function of network order $N(t)$, with the exponent  is $\frac{\ln(1+\sqrt{3})}{\ln(5)}$.In contrast to many recently studied network models mimicking real-life systems in nature and society \cite{Ref13,Ref14,Ref15,Ref16}, Sierpinski pentagons are not small worlds .

\section{Conclusion}
\label{sec:4}
In this paper,we have obtained rigorously  solution for the diameter and approximate solution for average path length ,both diameter and APL  of Sierpinski pentagons grow approximately as a power-law function of network order $N(t)$.Although the solution for APL  is approximate,it is trusted because we have calculated all items of APL accurately except for the compensation( $\Delta_{t}$) of  total distances between non-adjacent branches( $\Lambda_t^{1,3}$),which is obtained  approximately by least-square curve fitting.The compensation( $\Delta_{t}$) is only a small part of total distances between non-adjacent branches( $\Lambda_t^{1,3}$) and has little effect on  APL. Further more ,we use the data obtained by iteration to test our fitting results and find the relative error for  $\Delta_{t}$ is less than $10^{-7}$ which is acceptable.Hence the approximate solution for average path length is almost accurate.

\section*{Acknowledgment}
This research was supported by the National High Technology Research and Development Program("863"Program) of China under Grant No. 2009AA01Z439.



\begin{thebibliography}{}

\bibitem{Ref1}
R. Albert and A.-L. Barab\'asi,
       Rev. Mod. Phys. {\bf 74}, 47 (2002).
\bibitem{Ref2}
S.N. Dorogvtsev and J.F.F. Mendes ,  Adv. Phys.  {\bf51} ,1079(2002).
 \bibitem{Ref3}
M.E.J. Newman { SIAM Rev. }  {\bf45} 167(2003).
\bibitem{Ref4}
 A.-L. Barab\'{a}si, R. Albert ,{ Science} {\bf286} ,509(1999 ).
\bibitem{Ref5}
A.-L. Barab\'{a}si, R. Albert, H. Jeong, and G. Bianconi ,{ Science }  {\bf 287},  2115(2000).
\bibitem{Ref6}
Z. N. Oltvai, A.-L. Barab¨¢si, { Science }  {\bf 298},  763(2002).
\bibitem{Ref7}
M.E.J.Newman, { Phys. Rev. Lett.}  {\bf 89} , 208701(2002).
\bibitem{Ref8}
P. L.Krapivsky, and S. Redner,{ Phys. Rev. E }  {\bf 63}, 066123(2001).
\bibitem{Ref9}
 M. E. J. Newman, S. H. Strogatz, and D. J. Watts, { Phys.Rev. E }  {\bf 64}, 026118(2001).
\bibitem{Ref10}
Y.C.Zhang, Z.Z.Zhang, J.H.Guan, and S.G.Zhou.,{ J. Stat. Mech. }  {\bf 03}, P03013(2010).
\bibitem{Ref11}
I. Farkas, I. Derenyi, A-L. Barab¨¢si, T. Vicsek,{ Phys.Rev. E }  {\bf 64}, 026704(2001).
\bibitem{Ref12}
Z.Z.Zhang , Y.Qi, S.G.Zhou, Y.Lin, and J.H.Guan,{ Phys.Rev. E }  {\bf 80}, 016104(2009).
\bibitem{Ref48}
S.Jalan , G.M.Zhu , and B.W.Li ,{ Phys. Rev. E  }  {\bf 84}, 046107(2011) .
\bibitem{Ref49}
A.Yamamoto, S.Yamada, M.Okumura, and M.Machida,{ Phys. Rev. A }  {\bf 84}, 043642(2011).
\bibitem{Ref13}
R. Albert,H. Jeong,and A.-L. Barab\'{a}si ,{ Nature(London) }  {\bf401}, 130(1999).
\bibitem{Ref14}
A. Barratand, M. Weigt ,{ Eur. Phys. J. B }  {\bf13}, 547(2000).
\bibitem{Ref15}
 D.J. Watts, H. Strogatz ,{ Nature (London) }  {\bf 393}, 440(1998).
 \bibitem{Ref16}
 M. Barth\'{e}l\'{e}my and L. A. N. Amaral ,{ Phys. Rev. Lett.  }  {\bf82}, 5180(1999 ).
 \bibitem{EEstrada08}
E. Estrada, N. Hatano, Phys. Rev. E {\bf 77},  036111 (2008).  
\bibitem{EEstrada09}
E. Estrada, N. Hatano,Appl. Math. Comp. {\bf 214}  500-511(2009).
  \bibitem{Ref17}
 F. Chung and L.Lu, { Proc. Nat. Acad. Sci. }  {\bf99} , 15879(2002).
 \bibitem{Ref18}
 R. Cohen and S. Havlin, { Phys. Rev. Lett. }  {\bf90},  058701(2003 ).
 \bibitem{Ref19}
  C. Song , S. Havlin and H. A. Makse, { Nat. Phys. }  {\bf2} ,275(2006).
 \bibitem{Ref20}
  Z. Z. Zhang , S. G. Zhou and T. Zou, { Eur. Phys. J. B  }  {\bf56}, 259(2007).
 \bibitem{Ref21}
R. Pastor-Satorras and A. Vespignani ,{  Phys. Rev. Lett. }  {\bf86}, 3200(2001).
\bibitem{Ref22}
L. Ancel Meyers, M. E. J. Newman, M. Martin, and S. Schrag ,{  Emerging Infectious Diseases}  {\bf9}, 204(2001).
\bibitem{Ref23}
A. L. Lloyd and R. M. May ,{  Science }  {\bf292}, 1316(2001).
\bibitem{Ref24}
G. Yan , T. Zhou, B. Hu, Z. Q. Fu and B. H.Wang , { Phys. Rev. E } {\bf73} ,046108( 2006).
\bibitem{Ref25}
Z. Z. Zhang , F. Comellas, G. Fertin , A. Raspaud, L. L. Rong and S. G. Zhou , { J. Phys. A: Math. Theor.} {\bf41}, 035004(2008).
\bibitem{Ref26}
F. Comellas, A. Miralles ,{ J. Phys. A: Math. Theor. }  {\bf 42 }, 425001(2009).
\bibitem{Ref27}
Z.Z.Zhang, S. G.Zhou, T.Zou and  G. S.Chen, { J. Stat. Mech.} ,P09008(2008).
\bibitem{Ref28}
F. Jasch and A. Blumen ,{ Phys. Rev. E }  {\bf63}, 041108(2001).
\bibitem{Ref29}
M. F. Shlesinger ,{ Nature(London)}  {\bf443}, 281(2006).
\bibitem{Ref30}
M. E. J .Newman,  and D. J. Watts,{  Phys. Lett. A}  {\bf263}, 341(1999a).
\bibitem{Ref31}
B.Bollob\'{a}s and O. Riordan,{ Combinatorica }  {\bf24}, 5¨C34(2004).
\bibitem{Ref32}
Z.Z.Zhang ,L.C.Chen ,S. G.Zhou,  L.j.Fang,  J.h.Guan,  T.Zou,,{ Phys. Rev. E }  {\bf77} , 017102(2008).
\bibitem{Ref33}
Z.Z.Zhang ,L.C.Chen ,S. G.Zhou, L.j.Fang, S.G.Zhou,Y.C.Zhang,J.h.Guan,{ J. Stat. Mech. }  {\bf 02}, P02034(2009).
\bibitem{Ref34}
Z.Z.Zhang ,L. Yuan,S.Y.Gao,S.G.Zhou,J.h.Guan,{ J. Stat. Mech. },  P10022(2009).
\bibitem{Ref35}
B.Mandlebrot ,\emph{ The Fractal Geometry of Nature } (San Francisco: Freeman)(1982).
\bibitem{Ref36}
R. Engelking,{ Wiadomosci matematyczne }  {\bf 26(1)}, 18-24(1984).
\bibitem{Ref37}
L.J. Bentz, J. W. Turner, and J. J. Kozak,{ Phys. Rev. E }  {\bf 82}, 011137(2010).
\bibitem{Ref38}
S. H. Liu and A. J. Liu,{ Phys. Rev. B }  {\bf 32}, 4753 (1995).
\bibitem{Ref39}
Y.O. Hayase and T.Ohta,{ Phys. Rev. Lett. }  {\bf 81},  1726 (1998).
\bibitem{Ref40}
J.J.Kozak and V. Balakrishnan ,{ Phys. Rev. E } {\bf 65}, 021105(2002).
\bibitem{Ref47}
S.G.Ri, H.J.Ruan ,{ Journal of Mathematical Analysis and Applications} {\bf 380(1)}, 313-322(2011).
\bibitem{Ref41}
M. Fritsche, H. E. Roman, and M. Porto,{ Phys. Rev. E }  {\bf 76}, 061101(2007).
\bibitem{Ref42}
P.Y. Hsiao and P. Monceau,{ Phys. Rev. B  }  {\bf 65} ,184427(2002).
\bibitem{Ref43}
M. A. Bab, G. Fabricius, and E. V. Albano,{ Phys. Rev. E }  {\bf 71}, 036139(2005).
\bibitem{Ref44}
Y. Liu, Z. Hou, P. M. Hui, and W. Sritrakool,{  Phys. Rev. B}  {\bf 60} , 13444(1999).
\bibitem{Ref45}
A. Ordemann, M. Porto, and H. Ed. Roman,{ Phys. Rev. E }  {\bf 65}, 021107(2002).
\bibitem{Ref46}
Richard.A.Brualdi , \emph{Introductory Combinatorics },(Third Edition), (Pearson Education,Inc,1999).
\end{thebibliography}
\end{document}